\documentclass[pra,twocolumn,amsmath,showpacs]{revtex4}
\usepackage{graphicx}

\DeclareMathOperator{\Li}{Li}
\providecommand{\vect}[1]{\boldsymbol{#1}}

\begin{document}

\title{Shear viscosity and spin diffusion in a two-dimensional Fermi gas}

\author{Tilman Enss}
\affiliation{Physik Department, Technische Universit\"at M\"unchen,
  James-Franck-Str., 85747 Garching, Germany}
\author{Carolin K\"uppersbusch}
\author{Lars Fritz}
\affiliation{Institut f\"ur Theoretische Physik, Universit\"at zu
  K\"oln, Z\"ulpicher Stra\ss e 77, 50937 K\"oln, Germany}

\date{\today}

\begin{abstract}
  We investigate the temperature dependence of the shear viscosity and
  spin diffusion in a two-dimensional Fermi gas with contact
  interactions, as realized in ultra-cold atomic gases.  We describe
  the transport coefficients in terms of a Boltzmann equation and
  present a full numerical solution for the degenerate gas.  In
  contrast to previous works we take the medium effects due to finite
  density fully into account.  This effect reduces the viscosity to
  entropy ratio, $\eta/s$, by a factor of three, and similarly for
  spin diffusion.  The trap averaged viscosity agrees well with recent
  measurements by Vogt \emph{et al.}\ [Phys.\ Rev.\ Lett.\ \textbf{108},
  070404 (2012)].
\end{abstract}


\maketitle


\section{Introduction}

Ultra-cold atoms have emerged as a versatile system to study quantum
effects in strongly interacting fermionic and bosonic many-body
systems with excellent control over the Hamiltonian parameters
\cite{bloch2008}.  Transport properties provide particularly valuable
probes which can reveal the nature and strength of the effective
interaction between particles.  The shear viscosity $\eta$, e.g.,
measures the internal friction in a quantum fluid, which is lowest for
strongly interacting systems. For certain relativistic gauge theories
the ratio of the shear viscosity to the entropy density $s$ has been
computed using the AdS/CFT correspondence and takes the value
$(\eta/s)_\text{min} = \hbar/(4\pi k_B)$ \cite{policastro2001}.  It
has been conjectured that this value provides a lower bound also for a
wider class of relativistic field theories \cite{kovtun2005}, and
quantum fluids which saturate this bound are denoted as ``perfect
fluids'' \cite{schaefer2009}.  Subsequently quantum fluids ranging
from (non-relativistic) ultra-cold atoms to (relativistic) quark-gluon
plasmas have been investigated in the search for a perfect fluid which
saturates this bound \cite{schaefer2009}.  In the solid state context,
the viscosity of 2d graphene layers has been shown to decrease
logarithmically with increasing temperature \cite{mueller2009} coming
reasonably close to the lower limit.  Another example is the viscosity
of the unitary Fermi gas in three dimensions (3d) which has been
measured recently \cite{turlapov2008, cao2011, cao2011njp} and comes
rather close to the hypothetical bound for temperatures below the
Fermi temperature.  This is in agreement with calculations based on
kinetic theory for low \cite{landau1949, pethick1975, rupak2007} and
high temperatures \cite{massignan2005, bruun2005, riedl2008,
  enss2012}.  These calculations have been confirmed and refined in
approaches based on the Kubo formula with self-energy \cite{bruun2007}
and full vertex corrections \cite{enss2011} and recently also in the
form of a Quantum Monte Carlo simulation \cite{wlazlowski2012}.  A
similar lower bound is also seen in the spin diffusion coefficient $D$
which has a minimum close to the quantum limit $\sim \hbar/m$
\cite{sommer2011a, sommer2011b} again in good agreement with
calculations based on kinetic theory \cite{sommer2011a, bruun2011a}.

Recently, interacting ultra-cold gases have been realized in two
dimensions (2d) where quantum and interaction effects are even
stronger than in three dimensions \cite{froehlich2011, schmidt2012,
  dyke2011, sommer2011evolution, koschorreck2012}.  Measurements for a
trapped two-component 2d Fermi gas with strong interactions have found
the viscosity to decrease with decreasing temperature and increasing
interaction strength \cite{vogt2012}.

In this work we compute the shear viscosity $\eta$ and the spin
diffusion coefficient $D$ of an interacting two-component 2d Fermi gas
within kinetic theory.  Previous studies have investigated transport
without medium effects on the scattering cross section
\cite{schaefer2012, bruun2012, wu2012} and found a minimum value
$\eta/s \approx 20 (\eta/s)_\text{min}$ \cite{schaefer2012,
  bruun2012}.  We now include medium scattering, which is known to
strongly influence the dynamical properties \cite{bruun2005,
  riedl2008, schmidt2011, schmidt2012, enss2012}, and find that it
substantially lowers the viscosity by a factor of about three already
above $T_c$.  For the spin diffusion coefficient we find a similar
reduction.

The organization of the paper is as follows: In section
\ref{sec:model} we introduce the model Hamiltonian, the $T$-matrix in
medium, and then derive the quantum kinetic equations in section
\ref{sec:kin}.  A discussion of the zero mode in spin diffusion is
found in section \ref{sec:kin:spin}.  Readers familiar with the
Boltzmann approach may skip ahead directly to the results which are
presented in section \ref{sec:results}.  We close with a comparison to
experiment in section \ref{sec:exp} and conclude in section
\ref{sec:conclusion}.


\section{The model}
\label{sec:model}

We consider two species $\sigma=\;\uparrow, \downarrow$ of fermionic
atoms in two dimensions, which are described by the grand canonical
Hamiltonian
\begin{equation*}
  H = \sum_{\vect{k}\sigma} (\varepsilon_{\vect{k}\sigma} - \mu_\sigma)
  c_{\vect{k}\sigma}^\dagger c_{\vect{k}\sigma} + \frac{g_0}{V} \sum_{\vect{k}\vect{k}'\vect{q}}
  c_{\vect{k}\uparrow}^\dagger c_{\vect{k}'\downarrow}^\dagger
  c_{\vect{k}'-\vect{q}\downarrow} c_{\vect{k}+\vect{q}\uparrow},
\end{equation*}
with the free single-particle dispersion $\varepsilon_{\vect{k}\sigma} =
\vect{k}^2/2m_\sigma$ ($\hbar\equiv 1$), spin-dependent chemical potential
$\mu_\sigma$ and area $V$.  The model is formulated for the general
case of a heteronuclear mixture with different values for
$m_\uparrow$, $m_\downarrow$, $\mu_\uparrow$, and $\mu_\downarrow$,
however all numerical calculations are carried out for the balanced
case $\mu_\uparrow = \mu_\downarrow = \mu$ for equal masses
$m=m_\uparrow = m_\downarrow$ in view of the experiment
\cite{vogt2012}.  At ultra-cold temperatures the attractive $s$-wave
contact interaction $g_0$ acts only between different species due to
the Pauli principle.  The two-body scattering between a single
$\uparrow$ and $\downarrow$ fermion is given by the exact two-body
$T$-matrix \cite{adhikari1986, randeria1989}
\begin{equation}
  \label{eq:T0}
  \mathcal T_0(E) = \frac{2\pi/m_r}{\ln(\varepsilon_B/E) + i\pi}
\end{equation}
in terms of the reduced mass $m_r^{-1} = m_\uparrow^{-1} +
m_\downarrow^{-1}$.  The pole at $E=-\varepsilon_B < 0$ corresponds to
the two-body bound state, and the binding energy $\varepsilon_B =
1/(2m_ra_\text{2D}^2)$ defines the 2d scattering length $a_\text{2D}$.
This bound state is always present in an attractive 2d Fermi gas
\cite{landau1981, randeria1989}.  The vacuum scattering amplitude for
two particles with momenta $\vect{k}$ and $-\vect{k}$ in the center-of-mass
frame is then given by \cite{bloch2008} $f(k=|\vect{k}|) =
2m_r\mathcal T_0(k^2/2m_r) = 4\pi/[\ln(1/k^2a_\text{2D}^2)+i\pi]$.
The scattering amplitude depends logarithmically on energy in both the
low- and the high-energy limits: this is due to anomalous
(logarithmic) quantum corrections to the classically scale invariant
contact interaction \cite{hofmann2011, langmack2012}.

At finite density the two-particle scattering in the presence of the
medium is described by the many-body $T$-matrix $\mathcal
T(\vect{q},\omega)$.  It can be calculated from the solution of the
Bethe-Salpeter equation for the ladder approximation of repeated
particle-particle scattering \cite{nozieres1985, engelbrecht1990,
  engelbrecht1992}, and in the general case of spin imbalance it is
given by \cite{schmidt2012}
\begin{multline}
  \label{eq:Tinv}
  \mathcal T^{-1}(\vect{q},\omega) =
  \mathcal T_0^{-1}(\omega+i0+\mu_\uparrow+ \mu_\downarrow - \omega_{\vect{q}}) \\
  + \int \frac{d^2k}{(2\pi)^2}\,
  \frac{f_\uparrow^0(\vect{k}) +
    f_\downarrow^0(\vect{k}+\vect{q})}
  {\omega + i0 + \mu_\uparrow + \mu_\downarrow - \varepsilon_{\vect{k}\uparrow}
    - \varepsilon_{\vect{k}+\vect{q}\downarrow}}
\end{multline}
with the Fermi-Dirac distribution 
\begin{eqnarray}
  \label{eq:FD}
  f^0_\sigma(\vect{k}) = \frac{1}{e^{\beta(\varepsilon_{\vect{k}\sigma}-\mu_\sigma)}+1},
\end{eqnarray}
$\beta=1/(k_B T)$ and $\omega_{\vect{q}} = q^2/(8m_r)$.  While this
integral is known analytically at $T=0$ \cite{schmidt2012}, at finite
temperature we can perform only the angular average analytically but
have to compute the radial integral numerically. Compared to the case
with the bare $T$-matrix this increases the numerical effort in
solving the Boltzmann equation considerably.


\section{Transport properties from the kinetic approach}
\label{sec:kin}

We use the kinetic approach to derive the transport coefficients in
our system. This approach is valid provided quantum interference
effects are negligible and deviations from well-defined quasiparticles
are small, which we assume in the following. This assumption is
questionable for temperatures well below the Fermi temperature $T_F$
and the results should be compared to calculations within a formalism
which does not require the quasiparticle picture to be valid
\cite{enss2011, wlazlowski2012}.

The Boltzmann equation reads
\begin{eqnarray}
  \label{eq:Boltz}
  [\partial_t + \vect{v} \partial_{\vect{x}}
  +\vect{F}_\text{ext} \partial_{\vect{k}}]
  f_{\sigma}(\vect{k})
  = -I_\text{coll}[f_\sigma,f_{-\sigma}]\;, 
\end{eqnarray} 
which is an integro-differential equation for the quasiparticle
distribution function $f_{\uparrow,\downarrow}(\vect{k})$. The left-hand
side accounts for perturbations driving the system away from the
equilibrium situation, while the right-hand side accounts for
collisions between quasiparticles.

\subsection{General formalism: variational approach}
\label{sec:va}

The approach we take is standard but we present it such that
generalizations are possible in a straightforward manner. An excellent
account of this approach has been given in Refs.~\cite{ziman,
  smith1989, arnold} among others.  The left-hand side of
Eq.~\eqref{eq:Boltz} consists of three independent differential
operators and is henceforth referred to as the driving term, owing to
the fact that they drive the system away from equilibrium. The
individual terms describe temporal variations ($\partial_t$), spatial
variations ($\partial_{\vect{x}}$), as well as external forces
($\partial_{\vect{k}}$), while the right hand side describes
collisions due to interactions (or in other systems also disorder) and
consequently is called the collision integral. One can solve for the
non-equilibrium distribution function in the linear response regime
assuming that the deviation from the equilibrium distribution function
can be obtained in an expansion in the perturbation. This
schematically assumes the form
\begin{eqnarray}
  \label{eq:exp}
  f_{\sigma}({\vect{k}})=f^0_{\sigma}({\vect{k}})+\frac{1}{T}f^0_{\sigma}({\vect{k}})
  \left(1-f^0_{\sigma}({\vect{k}})\right)f^1_{\sigma}({\vect{k}})
\end{eqnarray}
for fermions where $f^0_\sigma ({\vect{k}})$ is the Fermi-Dirac
distribution, see Eq.~\eqref{eq:FD}, and $f^1_\sigma$ is linear in the
perturbation and otherwise a generic function (this is true for any
type of perturbation considered here). The factor
$\frac{1}{T}f^0_{\sigma}({\vect{k}})
\left(1-f^0_{\sigma}({\vect{k}})\right)$ is introduced for later
convenience. In this limit it is consistent to approximate the
collision integral by
\begin{eqnarray}
  I_\text{coll}[f_{\sigma},f_{-\sigma} ]
  &=& C [f_{\sigma}^1,f_{-\sigma}^1 ]
  +\mathcal{O}(f^2_\sigma,f^2_{-\sigma}) \nonumber \\
  &\approx& C [f_{\sigma}^1,f_{-\sigma}^1 ]
\end{eqnarray}
with 
\begin{widetext}
\begin{eqnarray}
  \label{eq:coll}
  && C[f_{\sigma}^1,f_{-\sigma}^1]
  =\frac{1}{T}\int_{\vect{k}_1,\vect{q}} \delta
  \bigl(\varepsilon_{\vect{k}\sigma}+\varepsilon_{\vect{k}_1-\sigma}
  -\varepsilon_{\vect{k}+\vect{q}\sigma}-\epsilon_{\vect{k}_1-\vect{q}-\sigma}
  \bigr)
  \Bigl|  \mathcal T \left( \vect{k}+\vect{k}_1,
    \varepsilon_{{\vect{k}}\sigma}+\varepsilon_{{\vect{k}_1}-\sigma}
    -\mu_\sigma-\mu_{-\sigma} \right) \Bigr|^2 \nonumber \\
  &&\times \left[ f^0_{\sigma}\left( {\vect{k}}\right) f^0_{-\sigma}\left(
      {\vect{k}}_1\right)\left(1- f^0_{\sigma}\left(
        {\vect{k}}+{\vect{q}}\right) \right)\left(1- f^0_{-\sigma}\left(
        {\vect{k}}_1-{\vect{q}}\right) \right) \right]
  \left[f^1_{\sigma}\left( {\vect{k}}\right)+f^1_{-\sigma}\left(
      \vect{k}_1\right)-f^1_{\sigma}\left(\vect{k}+\vect{q}\right)
    -f^1_{-\sigma}\left(\vect{k}_1-\vect{q}\right) \right] 
\end{eqnarray}
\end{widetext}
where $\int_{\vect{k}}=\int \frac{d^2k}{(2\pi)^2}$.

We are interested in the stationary solution, {\it i.e.}, $\partial_t
f_\sigma=0$ on the left-hand side of Eq.~\eqref{eq:Boltz}. To linear
order in the perturbation one can replace $f_\sigma\to f_\sigma^0$ on
the left-hand side and write 
\begin{eqnarray}
  D_{\alpha} f^0_{\sigma}=-C[f^1_{\sigma},f^1_{-\sigma}]\;,
\end{eqnarray}
where $D_\alpha$ in the most generic case is a tensor differential
operator acting on $f^0_{\sigma}$ and $\alpha$ labels the perturbation
we consider. In general we have
\begin{eqnarray}
  \mathcal{D}_{\alpha}^\sigma \equiv D_{\alpha} f^0_{\sigma}
  =-\frac{1}{T}f^0_\sigma \left(1-f^0_\sigma \right) I^{ij}_\sigma F^{ij}_\sigma
\end{eqnarray}
where we use the Einstein summation convention. At this point we have
introduced $F^{ij}_\sigma$ as a generalized force field and
$I^{ij}_\sigma$ as a generalized projection. For reasons of a concise
presentation we assume from now on that we can absorb the
spin-dependence of $F^{ij}_\sigma$ into the factor $I^{ij}_\sigma$ and
work with $F^{ij}$ only which acts in the same way on both spin
species. For concreteness, in the case of an electrical conductivity
we have $F^{ij}_\sigma=E^i \delta_{ij}$ and
$I_\sigma^{ij}=ev^i_{{\vect{k}},\sigma}\delta_{ij}$.  This general
form also dictates the form of the ansatz for $f^1_\sigma$, which we
choose as
\begin{eqnarray}
  \label{eq:modes}
  f^1_\sigma({\vect{k}})=F^{ij}_\sigma \chi^{ij}_\sigma ({\vect{k}})
  =F^{ij}_\sigma I^{ij}_\sigma g_\sigma (k) \;.
\end{eqnarray}
The generalized current then reads
\begin{eqnarray}
  \label{eq:j}
  j^{ij}&=&\sum_\sigma \int_{{\vect{k}}} I^{ij}_\sigma f_\sigma ({\vect{k}}) \nonumber \\
  &=&\sum_\sigma \frac{1}{T} \int_{\vect{k}} f^0_\sigma (1-f^0_\sigma)
  I^{ij}_\sigma F^{kl}_\sigma \chi^{kl}_\sigma \nonumber \\
  &=&-\sum_\sigma \int_{{\vect{k}}}\chi^{ij}_\sigma \mathcal{D}_{\alpha}^\sigma \nonumber\\
  &=&-\langle \chi^{ij} | \mathcal{D}_{\alpha}\rangle =-\mathcal{S}[\chi^{ij}]\, F^{ij}
\end{eqnarray}
where $|\chi^{ij} \rangle=(\chi^{ij}_{\uparrow},
\chi^{ij}_{\downarrow})$ is a spinor and the components are themselves
vectors in function space. In the last line we have introduced a
scalar product. Using this definition of a scalar product we can also
define
\begin{eqnarray}
  \mathcal{C}[\chi^{ij}]=\frac{1}{2} \langle \chi^{ij} |C | \chi^{ij}
  \rangle F^{ij}  \;.
\end{eqnarray}
We can now introduce a functional
\begin{eqnarray}
  \mathcal{Q}[\chi^{ij}]=\mathcal{S}[\chi^{ij}]+\mathcal{C}[\chi^{ij}]
\end{eqnarray}
whose extremum in function space 
\begin{eqnarray}\label{eq:M}
  \partial_{\chi_\sigma} \mathcal Q [\chi^{ij}]
  \bigg|_{\chi_{\sigma}^{ij,\text{max}}} = 0
\end{eqnarray}
can be shown to lead to the Boltzmann equation for the respective
species. Conversely, the Boltzmann equation implies that the current
reads
\begin{eqnarray}
  \label{eq:jQ}
  j^{ij}&=&-\mathcal{S}[\chi_\sigma^{ij,\text{max}}]\, F^{ij}
  =2\mathcal{C}[\chi_\sigma^{ij,\text{max}}]\, F^{ij}\nonumber \\
  &=&-2 \mathcal{Q}[\chi_\sigma^{ij,\text{max}}]\,  F^{ij}\;. 
\end{eqnarray} 
The proper strategy to solve the Boltzmann equation is thus to
maximize the functional $\mathcal{Q}[\chi_\sigma^{ij}]$ for
$\chi_\sigma^{ij}=I_\sigma^{ij} g_\sigma(k)$ by varying $g_\sigma(k)$.
This is done by identifying the physically most relevant modes
$g_{n\sigma}(k)$ and writing $g_\sigma(k)$ as an expansion with
respect to these modes:
\begin{eqnarray}\label{eq:VA}
  g_\sigma(k)=\sum_n \lambda_n g_{n\sigma} (k)\;.
\end{eqnarray}
Maximizing $\mathcal{Q}[g_\sigma(k)]$ with respect to the expansion
coefficients $\lambda_n$ leads to a matrix equation for $\lambda_n$
which can be solved by matrix inversion.  Usually the most relevant
modes are the slow modes which are related to almost conserved
quantities whose relaxation is described by the collision kernel.

\subsection{The shear viscosity within Boltzmann theory}

We consider a two-component Fermi gas in its most general form,
allowing for different chemical potentials for the two species, {\it
  i.e.}, $\mu_{\uparrow}$ and $\mu_{\downarrow}$ and a species
dependent mass $m_\sigma$. We are concerned with a system without
external forces, {\it i.e.}, ${\vect{F}}_\text{ext}={\vect{0}}$ in its
stationary state $\partial_t f_\sigma=0$. We assume a uniform flow in
x-direction and a velocity gradient in y-direction, {\it i.e.},
${\vect{u}}=(u(y),0)$ which leads us to analyze the Boltzmann equation
according to
\begin{eqnarray}
  \label{eq:flow}
  {\vect{v}} \partial_{\vect{x}} f_{\sigma}({\vect{k}})
  =-I_\text{coll}[f_{\sigma},f_{-\sigma}]\;.
\end{eqnarray}
The collision term for the contact interaction in its linearized
version was introduced in Eq.~\eqref{eq:coll}. The driving term reads
\begin{eqnarray}
  \mathcal{D}^{\sigma}_{\eta}
  &=&-\frac{k_xk_y}{m_\sigma T} \frac{\partial u}{\partial y}
  f^0_{\sigma}({\vect{k}})\left(1-f^0_{\sigma}({\vect{k}})\right) \;.
\end{eqnarray}
Following the logic of section \ref{sec:va} we define more generally
\begin{eqnarray}
  F^{ij}_\sigma&=&\partial_iu_j+\partial_ju_i
  - \frac{2}{d}\delta_{ij}\partial_lu_l \nonumber \\
  I^{ij}_\sigma&=& v_{{\vect{k}},\sigma}^i k^j
\end{eqnarray}
with $u_i$ being the components of the flow velocity of the fluid.
The generalized current is the viscous part of the stress tensor
describing hydrodynamics in two spatial dimensions,
\begin{eqnarray}
 \label{eq:viscous stress tensor}
 j^{ij}=-\eta F^{ij}-\zeta \delta_{ij}\partial_lu_l
\end{eqnarray}
where $\eta$ is the shear viscosity and $\zeta$ the bulk viscosity.

Combining \eqref{eq:viscous stress tensor} and \eqref{eq:j} one obtains
\begin{eqnarray}
 \eta = \mathcal S[\chi^{ij}]
\end{eqnarray}
for the exact solution $|\chi^{ij} \rangle$.  The variational
principle provides us with a lower bound.  If we make an ansatz
$|\chi^{\text{ansatz}} \rangle$ using a \emph{finite} function set
$g_{n\sigma}(k)$ this implies \cite{smith1989}
\begin{eqnarray}
 \eta \geq  \mathcal S[\chi^{\text{ansatz}}]
 \bigg|_{\chi^{\text{ansatz}}=\chi^{\text{ansatz}}_{\text{max}}}
\end{eqnarray}
where $| \chi^{\text{ansatz}}_{\text{max}} \rangle$ corresponds to the
optimal choice for a finite number of the parameters $\lambda_n$
introduced in Eq.~\eqref{eq:VA} which maximizes Eq.~\eqref{eq:M}.  In
the case of the viscosity there is no conserved quantity which is
excited. We found that, just as in the three-dimensional case
\cite{bruun2007}, the choice for the modes Eq.~\eqref{eq:modes}
\begin{eqnarray}\label{eq:vmode}
g_\sigma(k)=1
\end{eqnarray}
yields results which are very close to the exact result. We have
checked this statement for different sets of modes, for instance
$g_{n\sigma}(k)=k^n$ for $n=0,\dotsc,N$ up to $N=10$ as well as
Chebyshev polynomials up to the same order and have found no
pronounced differences.

\subsection{Spin diffusion within Boltzmann theory}
\label{sec:kin:spin}

Spin diffusion in a metal describes the response of a system of
fermions to a gradient in a magnetic field. In our setup this
translates to the two fermion species responding to gradients in
chemical potentials, which are opposite for the two species. Again we
discuss the most generic situation, which is that there are two
species of fermions with different chemical potentials
$\mu_{\uparrow}\neq \mu_{\downarrow}$ and different atomic masses
$m_{\uparrow}\neq m_\downarrow$. We assume there is a chemical
potential for the individual atoms $\mu_\sigma+{\vect{r}}\cdot \nabla
\mu_{\sigma}$. The distribution function is accordingly driven out of
equilibrium by
\begin{eqnarray}
  \mathcal{D}_s^\sigma= -\frac{{\vect{k}}\cdot \nabla \mu_\sigma}{m_\sigma T}
  f^0_\sigma \left (1-f^0_\sigma \right)\;.
\end{eqnarray}
In the following we assume that the absolute value of the gradient is
the same for both species but counteracts, $\nabla \mu_{\sigma}=\sigma
\nabla \mu$, such that
\begin{eqnarray}
  \mathcal{D}_s^\sigma= -\sigma \frac{{\vect{k}} \cdot \nabla
    \mu}{m_\sigma T}   f^0_\sigma \left (1-f^0_\sigma \right)\;. 
\end{eqnarray}
Again, we identify the generalized force and projector
\begin{eqnarray}
  F^{ij}&=&\partial_i \mu \delta_{ij} \nonumber \\
  I^{ij}_\sigma&=& \sigma v_{\vect{k},\sigma}^i \delta_{ij} \;.
\end{eqnarray}
The spin conductivity is again bounded from below by
\begin{eqnarray}
  \sigma_s\geq \mathcal S[\chi^{\text{ansatz}}]
  \bigg|_{\chi^{\text{ansatz}}=\chi^{\text{ansatz}}_{\text{max}}}
\end{eqnarray}
and we can deduce the spin diffusion coefficient $D$ via
\begin{eqnarray}
  D=\frac{\sigma_s}{\chi_s}
\end{eqnarray}
with the spin susceptibility of the free Fermi gas
\begin{eqnarray}
  \chi_s=\frac{m_\uparrow f_\uparrow^0(k=0) + m_\downarrow f_\downarrow^0(k=0)}{2\pi}\;.
\end{eqnarray}
In the case of the viscosity the driving term does not couple to a
conserved quantity such as the total energy or the momentum.
Consequently, the variational approach can be employed with relatively
little care and very few modes suffice to solve the problem
essentially exactly.  In the case of the spin diffusion this ceases to
be true and the driving term in general does not decouple from the
momentum mode.  The momentum mode corresponds to the choice
\begin{eqnarray}
  g_\sigma=\sigma m_\sigma
\end{eqnarray}
and if we calculate the overlap of the momentum mode with the driving
term within this variational ansatz it reads
\begin{eqnarray*}
  \langle \chi | \mathcal{D}_s \rangle=\frac{T}{\pi}  
  \Bigl(m_\uparrow \ln (1+e^{\beta\mu_\uparrow} )
    - m_{\downarrow} \ln (1+e^{\beta\mu_\downarrow} )
  \Bigr) \;. \nonumber \\ 
\end{eqnarray*}
This is zero if $\mu_\uparrow=\mu_\downarrow=\mu$ and
$m_\uparrow=m_\downarrow=m$, meaning the momentum mode is not
excited. If these conditions do not hold the momentum mode is excited
and it cannot be relaxed. This formally leads to an infinite spin
conductivity $\sigma_s$.  In metals the standard situation is
spin balance with a finite spin conductivity, as has been discussed
recently in the context of graphene~\cite{mueller2011}. In the
experiments under discussion two clouds of different spin species are
prepared to collide in the center of the trap. If the two clouds are
equal in number of particles and masses the unified cloud will reside
in the center of the trap. One could excite the zero mode if one
prepared different densities and/or different masses for the different
spin species. The zero mode of the spin diffusion then has a very
simple and intuitive physical meaning and it corresponds to a center
of mass motion.

In our concrete setup in a balanced system we work with the choice
\begin{eqnarray}
  \label{eq:sdmode}
  g_\sigma= m
\end{eqnarray}
which is not a zero mode of the collision integral and has finite
overlap with the driving term. We have again checked more generic mode
choices and found this to provide an excellent variational ansatz.


\section{Results}
\label{sec:results}

We have obtained the viscosity and spin diffusion from the variational
approach using the variational ansatz functions introduced in
Eq.~\eqref{eq:vmode} and Eq.~\eqref{eq:sdmode}.  The transport
coefficients are normalized by the respective thermodynamic quantities
density, pressure, and entropy density, and for consistency they all
have to be computed at the same level of approximation.  A definite
prescription is provided by the large-$N$ expansion \cite{enss2012},
which interpolates between free fermions ($N=\infty$) and the physical
case of interacting fermions ($N=1$): to leading order in $1/N$, the
collision integral with the full medium scattering $T$-matrix is
consistent with using the density and pressure of the free Fermi gas.
Specifically, the density of a free balanced 2d Fermi gas is
\begin{align}
  \label{eq:freedens}
  n \lambda_T^2 = 2\ln(1+z) = 2/\theta
\end{align}
with thermal length $\lambda_T = \sqrt{2\pi/mk_BT}$ and fugacity
$z=\exp(\beta\mu)=\exp(1/\theta)-1$ in terms of the reduced
temperature $\theta = T/T_F$.  The pressure is expressed by the
polylogarithm $\Li_s(z)$ as
\begin{align}
  \label{eq:freepress}
  P = -nk_BT\theta\Li_2(1-e^{1/\theta}) 
\end{align}
and the internal energy density $\varepsilon = E/V = P$ equals the
pressure by scale invariance.  The entropy density
\begin{align*}
  s & = \frac{\varepsilon + P - \mu n}{T}
  = nk_B \bigl\{ -2\theta\Li_2(1-e^{1/\theta}) - \ln(e^{1/\theta}-1) \bigr\}
\end{align*}
becomes in the high temperature classical limit $\theta\to\infty$
\begin{align*}
  s = nk_B \bigl\{ 2+\ln\theta + \mathcal O(\theta^{-2}) \bigr\}.
\end{align*}
Henceforth we will set $k_B=1$.

\subsection{Viscosity}

\begin{figure}[ht]
  \includegraphics[width=\linewidth]{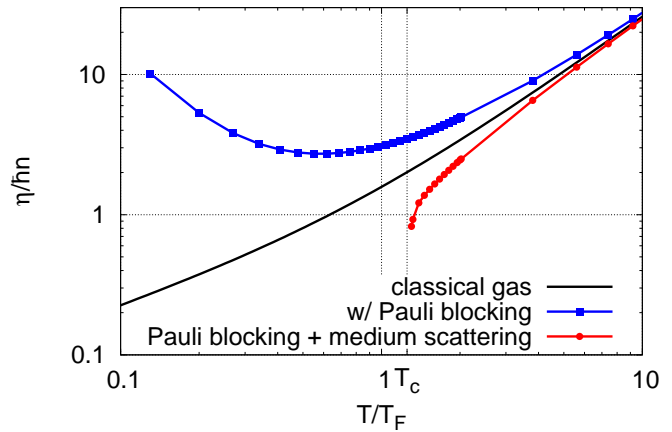}
  \caption{(color online) Shear viscosity $\alpha = \eta/n$ with and
    without medium effects, at strong interaction $\varepsilon_B/
    \varepsilon_F = 2$.  While Pauli blocking (squares) increases the
    viscosity with respect to the classical gas (solid line), medium
    scattering (circles) substantially lowers the minimum as $T_c$ is
    approached from above.}
  \label{fig:etan}
\end{figure}

We compute the viscosity of the strongly interacting 2d Fermi gas with
full medium effects.  The case with Pauli blocking and the bare vacuum
scattering cross section, including the limits of high and low
temperature, has been discussed in Refs.~\cite{schaefer2012,
  bruun2012, wu2012}.  Our main finding is that the medium increases
scattering for strong interaction and thereby substantially lowers the
transport coefficients, see Fig.~\ref{fig:etan}.  For vacuum
scattering (squares) the system always appears to be in the normal
Fermi liquid phase and the upturn of the viscosity for low
temperatures is due to Pauli blocking.  With medium scattering the
viscosity decreases down to a finite temperature $T_c$ where the
medium $T$-matrix acquires a pole, $\mathcal T^{-1}(q=0,\omega=0) = 0$
(Thouless criterion).  Below $T_c$ this pole would formally lead to a
diverging collision integral $C$ and $\eta\to0$ in this approximation.
A calculation of the viscosity in the superfluid $B$ phase of $^3$He for
$T<T_c$ found that Pauli blocking and enhanced scattering cancel
precisely and $\eta$ approaches a finite value for $T\to 0$
\cite{pethick1975}.

\begin{figure}[ht]
  \includegraphics[width=\linewidth]{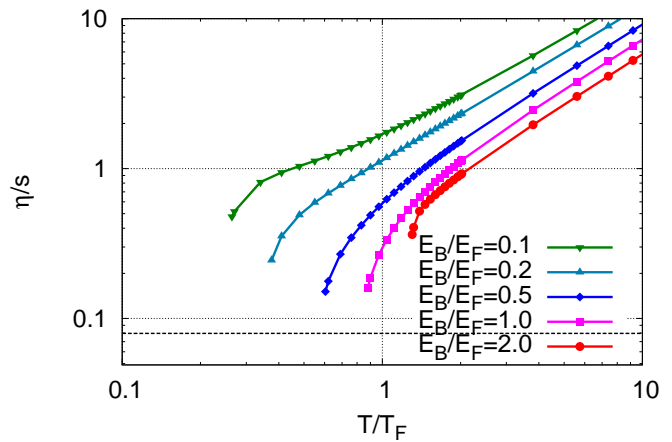}
  \caption{(color online) Viscosity to entropy ratio $\eta/s$ with
    medium scattering above $T_c$ for different interaction strengths
    $\varepsilon_B/\varepsilon_F = 0.1,\; 0.2,\; 0.5,\; 1,\; 2$ (from
    top to bottom).  The dashed line indicates the bound $1/(4\pi)$.}
  \label{fig:etas}
\end{figure}

In Fig.~\ref{fig:etas} the ratio of the viscosity to entropy density
$\eta/s$ is compared for different values of the interaction strength.
As the binding energy $\varepsilon_B$ is lowered, $T_c$ as defined by
the Thouless criterion is shifted to lower temperatures, indicated by
the endpoints of the solid lines (the endpoints are at $T=1.04\,
T_c$).  As an estimate, the minimum for $\varepsilon_B/ \varepsilon_F
= 0.5$ is located at around $T/T_F=0.6$ at a value of $\eta/s = 0.15$,
only about twice the proposed string theory bound $\eta/s = 1/(4\pi)$.

\subsection{Spin diffusion}

\begin{figure}[ht]
  \includegraphics[width=\linewidth]{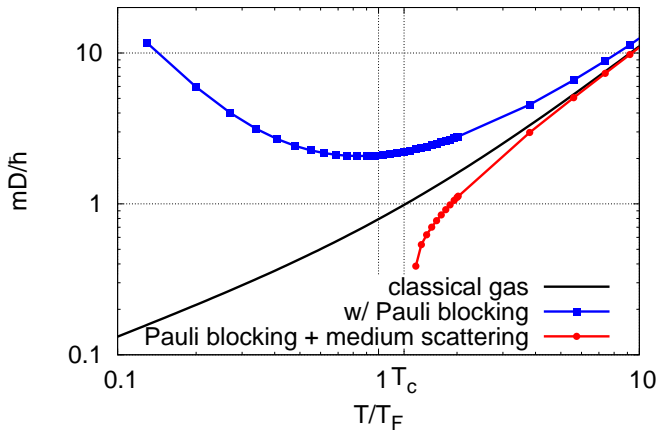}
  \caption{(color online) Spin diffusion coefficient $D$ in the
    high-temperature limit of a classical gas (solid line), including
    Pauli blocking (squares) and with the full medium scattering cross
    section (circles).}
  \label{fig:diff}
\end{figure}

Equivalently, we have carried out the analysis for the spin diffusion
coefficient $D$.  In the high-temperature limit \cite{bruun2012}
\begin{align}
  \label{eq:2}
  D & = \frac{Q\theta}{4\pi} &
  Q & = \pi^2 + \ln^2\left( \frac{3T}{2\varepsilon_B} \right)
\end{align}
the diffusion coefficient depends linearly on $\theta$ with
logarithmic corrections, see Fig.~\ref{fig:diff}.  Pauli blocking
(squares) increases diffusion, while the inclusion of medium effects
leads to a strong reduction of the diffusion coefficient $D$
(circles).

\section{Comparison to experiment}
\label{sec:exp}

\begin{figure}[ht]
  \includegraphics[width=\linewidth]{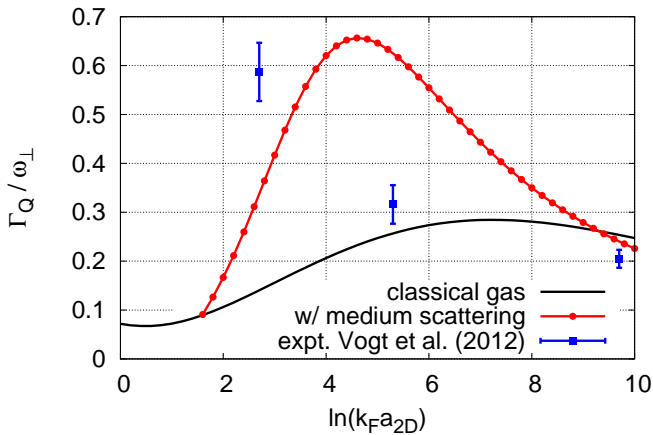}
  \caption{(color online) Quadrupole damping rate
    $\Gamma_Q/\omega_\perp$ vs.\ the interaction strength of the
    trapped gas at $T/T_F=0.3$ and $E_F/h = 6.4\,$kHz, with radial
    trapping frequency $\omega_\perp = 2\pi\times 125\,$Hz and
    $N=2620$ particles.}
  \label{fig:gamma}
\end{figure}

In order to compare our results for the balanced homogeneous 2d Fermi
gas with experiments in a trap we perform an average over the density
profile of the trap assuming the local density approximation to hold.
At high temperatures the density profile in the trap is
\cite{schaefer2012}
\begin{align}
  \label{eq:dens}
  n(r) = \frac{N}{\pi\sigma^2} e^{-r^2/\sigma^2}
\end{align}
with $\sigma^2=2T/(m\omega_\perp^2)$, radial trapping frequency
$\omega_\perp$ and total density $\int d^2r\, n(r) = N$.  The local
Fermi temperature is given in terms of the density as
\begin{align}
  T_F(r) = \frac{\pi}{m} n(r)
\end{align}
such that the local reduced temperature is
\begin{align}
  \theta(r) = \frac{T}{T_F(r)} = \frac{mT}{\pi n(r)}
\end{align}
and the local pressure of the free Fermi gas is, cf.\
\eqref{eq:freepress},
\begin{align}
  P(r) = -n(r) T \theta(r) \Li_2(1-e^{1/\theta(r)}).
\end{align}
The frequency dependent shear viscosity of the homogeneous system is
in kinetic theory \cite{braby2011, schaefer2012, enss2012}
\begin{align}
  \eta(\omega) = \frac{P\tau}{1+\omega^2\tau^2}
\end{align}
in accordance with the viscosity sum rule \cite{taylor2012}.  From the
dimensionless ratio $\eta(0)/n = \alpha(\theta)$ one obtains the
viscous scattering time
\begin{align}
  \label{eq:tau}
  \tau = \frac{\eta(0)}{P} = \frac{n}{P} \alpha(\theta).
\end{align}
The local viscosity can be defined in terms of the local reduced
temperature $\theta(r)$,
\begin{align}
  \eta(\omega,r) = \frac{n(r) \alpha(\theta(r))}
  {1+\omega^2[n(r)\alpha(\theta(r))/P(r)]^2}.
\end{align}
The spatial integral of the viscosity diverges at $\omega=0$ because
the dc viscosity is density independent in the outer regions of the
trap \cite{schaefer2012, bruun2012}.  In order to obtain a finite
integral the viscosity is evaluated at the quadrupole frequency
$\omega_Q = \sqrt 2 \omega_\perp$ \cite{vogt2012},
\begin{align}
  \langle\alpha\rangle = \frac{1}{N} \int d^2r\, \eta(\omega_Q,r) \,.
\end{align}
The global Fermi temperature $T_F = \sqrt N \omega_\perp$ allows us to
define a global reduced temperature $\Theta=T/T_F$, so that the trap
averaged viscosity can be written as
\begin{align}
  \label{eq:3}
  \langle\alpha(\Theta)\rangle = \frac 1N \int d^2r\, n(r)
  \frac{\alpha(\theta(r))}
  {1+\bigl(\frac{\omega_Q}{\omega_\perp}\bigr)^2
    \frac{\alpha^2(\theta(r))}{N \Theta^2 p^2(\theta(r))}}
\end{align}
with dimensionless pressure $p(\theta(r)) = P(r)/(n(r)T)$.  We can
change variables and integrate $\theta(r) = 2\Theta^2\dotsc\infty$,
\begin{align}
  \label{eq:1}
  \langle\alpha(\Theta)\rangle = 2\Theta^2 \int_{2\Theta^2}^\infty
  \frac{d\theta}{\theta^2} \frac{\alpha(\theta)} 
  {1+\bigl(\frac{\omega_Q}{\omega_\perp}\bigr)^2
    \frac{\alpha^2(\theta)}{N \Theta^2 p^2(\theta)}}
\end{align}
Finally, the quadrupole damping rate is \cite{vogt2012}
\begin{align}
  \frac{\Gamma_Q}{\omega_\perp}
  & = \frac{2\langle\alpha(\Theta)\rangle}{m\omega_\perp \langle
    r^2 \rangle}
  = \frac{\langle\alpha(\Theta)\rangle}{\sqrt N \Theta}
\end{align}
with $\langle r^2 \rangle = \sigma^2$ for the density profile in
Eq.~\eqref{eq:dens}.  In the high-temperature limit the integrals can
be solved analytically and yield \cite{schaefer2012}
\begin{align}
  \alpha(\theta) & = \frac{R \theta}{2\pi}, \qquad
  R = \pi^2 + \ln^2\left( \frac{5T}{2\varepsilon_B} \right) \\
  \langle\alpha(\Theta)\rangle & = \frac{R\Theta^2}{2\pi} \ln \left[
    1+\frac{\pi^2N}{2R^2\Theta^2} \right] \\
  \frac{\Gamma_Q}{\omega_\perp} & = \frac{R \Theta}{2\pi\sqrt N} \ln\left[
    1+\frac{\pi^2N}{2R^2\Theta^2} \right] 
\end{align}
where we have used $p(\theta)=1$ and $(\omega_Q/\omega_\perp)^2=2$.
In Fig.~\ref{fig:gamma} we show the quadrupole damping rate vs.\
interaction strength and compare with the experimental values
\cite{vogt2012}.  The effect of the medium scattering is most
pronounced at low temperature and strong interaction.  This leads to
strongly enhanced damping, and the peak height $\Gamma_Q/\omega_\perp
\sim 0.6$ agrees well with experiment, while previous theoretical
studies found lower peak values $\Gamma_Q/\omega_\perp \lesssim 0.4$
\cite{bruun2012, wu2012}.  Still, the peak position in our calculation
occurs at a larger interaction parameter than in the experiment.

\section{Conclusion}\label{sec:conclusion}

We have investigated the temperature dependence of the shear viscosity
and spin diffusion in a two-component Fermi gas in two dimensions with
contact interactions. We used the Boltzmann equation where in contrast
to former works we took the medium effect due to finite fermion
density into account. We show that the proper inclusion of this effect
leads to strong suppression of both transport quantities. Performing
the trap average we find that the inclusion of medium effects
quantitatively brings us rather close to the experimental
findings~\cite{vogt2012}.  It is an important question for the future
to confirm the result obtained within the Boltzmann framework with a
more refined calculation which does not rely on the validity of the
quasiparticle picture and possibly extends below $T_c$.


\paragraph*{Acknowledgements.}
We acknowledge discussions with M. K\"ohl, R. Schmidt and W. Zwerger,
and we thank M. K\"ohl for sending us data. LF acknowledges related
collaborations and many discussions with J. Lux, M. M\"uller,
J. Schmalian, and S. Sachdev. This work was supported by the
``Deutsche Forschungsgemeinschaft'' within SFB 608 (LF), the
Bonn-Cologne Graduate School (CK), and the Emmy-Noether Program Grant
No.\ FR 2627/3-1 (CK, LF).


\begin{thebibliography}{43}

\bibitem{bloch2008}
I.~Bloch, J.~Dalibard, and W.~Zwerger, Rev.\ Mod.\ Phys.\ \textbf{80}, 885
  (2008).

\bibitem{policastro2001}
G.~Policastro, D.~T. Son, and A.~O. Starinets, Phys.\ Rev.\ Lett.\ \textbf{87},
  081601 (2001).

\bibitem{kovtun2005}
P.~K. Kovtun, D.~T. Son, and A.~O. Starinets, Phys.\ Rev.\ Lett.\ \textbf{94},
  111601 (2005).

\bibitem{schaefer2009}
T.~Sch{\"a}fer and D.~Teaney, Rep.\ Prog.\ Phys.\ \textbf{72}, 126001 (2009).

\bibitem{mueller2009}
M.~M{\"u}ller, J.~Schmalian, and L.~Fritz, Phys.\ Rev.\ Lett.\ \textbf{103},
  025301 (2009).

\bibitem{turlapov2008}
A.~Turlapov, J.~Kinast, B.~Clancy, L.~Luo, J.~Joseph, and J.~E. Thomas, J. Low
  Temp.\ Phys.\ \textbf{150}, 567 (2008).

\bibitem{cao2011}
C.~Cao, E.~Elliott, J.~Joseph, H.~Wu, J.~Petricka, T.~Sch{\"a}fer, and J.~E.
  Thomas, Science \textbf{331}, 58 (2011).

\bibitem{cao2011njp}
C.~Cao, E.~Elliott, H.~Wu, and J.~E. Thomas, New J. Phys. \textbf{13}, 075007
  (2011).

\bibitem{landau1949}
L.~D. Landau and I.~M. Khalatnikov, Zh.\ Eksp.\ Teor.\ Fiz. \textbf{19},
  637 and 709 (1949), English translation in \textit{Collected Papers of L. D.
  Landau} (Pergamon Press, Oxford, 1965), p.~494.

\bibitem{pethick1975}
C.~J. Pethick, H.~Smith, and P.~Bhattacharyya, Phys.\ Rev.\ Lett.\ \textbf{34},
  643 (1975).

\bibitem{rupak2007}
G.~Rupak and T.~Sch{\"a}fer, Phys.\ Rev.~A \textbf{76}, 053607 (2007).

\bibitem{massignan2005}
P.~Massignan, G.~M. Bruun, and H.~Smith, Phys.\ Rev.~A \textbf{71}, 033607
  (2005).

\bibitem{bruun2005}
G.~M. Bruun and H.~Smith, Phys.\ Rev.~A \textbf{72}, 043605 (2005).

\bibitem{riedl2008}
S. Riedl, E. R. Sanchez Guajardo, C. Kohstall, A. Altmeyer,
M. J. Wright, J. Hecker Denschlag, R. Grimm, G. M. Bruun, and
H. Smith, Phys.\ Rev.~A \textbf{78}, 053609 (2008).

\bibitem{enss2012}
T.~Enss, Phys.\ Rev.~A \textbf{86}, 013616 (2012).

\bibitem{bruun2007}
G.~M. Bruun and H.~Smith, Phys.\ Rev.~A \textbf{75}, 043612 (2007).

\bibitem{enss2011}
T.~Enss, R.~Haussmann, and W.~Zwerger, Ann.\ Phys.\ (N.Y.) \textbf{326}, 770
  (2011).

\bibitem{wlazlowski2012}
G.~Wlaz{\l}owski, P.~Magierski, and J.~E. Drut, arXiv:1204.0270
[Phys. Rev. Lett. (to be published)].

\bibitem{sommer2011a}
A.~Sommer, M.~Ku, G.~Roati, and M.~W. Zwierlein, Nature (London) \textbf{472}, 201
  (2011).

\bibitem{sommer2011b}
A.~Sommer, M.~Ku, and M.~W. Zwierlein, New J. Phys.\ \textbf{13}, 055009 (2011).

\bibitem{bruun2011a}
G.~M. Bruun, New J. Phys.\ \textbf{13}, 035005 (2011).

\bibitem{froehlich2011}
B.~Fr{\"o}hlich, M.~Feld, E.~Vogt, M.~Koschorreck, W.~Zwerger, and M.~K{\"o}hl,
  Phys.\ Rev.\ Lett.\ \textbf{106}, 105301 (2011).

\bibitem{schmidt2012}
R.~Schmidt, T.~Enss, V.~Pietil{\"a}, and E.~Demler, Phys.\ Rev.~A \textbf{85},
  021602(R) (2012).

\bibitem{dyke2011}
P. Dyke, E. D. Kuhnle, S. Whitlock, H. Hu, M. Mark, S. Hoinka,
  M. Lingham, P. Hannaford, and C. J. Vale, Phys.\ Rev.\ Lett.\
  \textbf{106}, 105304 (2011).

\bibitem{sommer2011evolution}
A.~T. Sommer, L.~W. Cheuk, M.~J.~H.~Ku, W.~S. Bakr, and M.~W. Zwierlein,
  Phys.\ Rev.\ Lett.\ \textbf{108}, 045302 (2012).

\bibitem{koschorreck2012}
M.~Koschorreck, D.~Pertot, E.~Vogt, B.~Fr\"ohlich, M.~Feld, and M.~K\"ohl,
  Nature (London) \textbf{485}, 619 (2012).

\bibitem{vogt2012}
E.~Vogt, M.~Feld, B.~Fr\"ohlich, D.~Pertot, M.~Koschorreck, and M.~K\"ohl,
  Phys.\ Rev.\ Lett.\ \textbf{108}, 070404 (2012).

\bibitem{schaefer2012}
T.~Sch{\"a}fer, Phys.\ Rev.~A \textbf{85}, 033623 (2012).

\bibitem{bruun2012}
G.~M. Bruun, Phys.\ Rev.~A \textbf{85}, 013636 (2012).

\bibitem{wu2012}
L.~Wu and Y.~Zhang, Phys.\ Rev.~A \textbf{85}, 045601 (2012).

\bibitem{schmidt2011}
R.~Schmidt and T.~Enss, Phys.\ Rev.~A \textbf{83}, 063620 (2011).

\bibitem{adhikari1986}
S.~K. Adhikari, Am.\ J. Phys.\ \textbf{54}, 362 (1986).

\bibitem{randeria1989}
M.~Randeria, J.-M. Duan, and L.-Y. Shieh, Phys.\ Rev.\ Lett.\ \textbf{62}, 981
  (1989).

\bibitem{landau1981}
L.~D. Landau and E.~M. Lifshitz, \emph{Quantum Mechanics}
  (Butterworth-Heinemann, Oxford, 1981).

\bibitem{hofmann2011}
J.~Hofmann, Phys.\ Rev.~A \textbf{84}, 043603 (2011).

\bibitem{langmack2012}
C.~Langmack, M.~Barth, W.~Zwerger, and E.~Braaten, Phys.\ Rev.\ Lett.\
  \textbf{108}, 060402 (2012).

\bibitem{nozieres1985}
P.~Nozieres and S.~Schmitt-Rink, J. Low Temp.\ Phys.\ \textbf{59}, 195 (1985).

\bibitem{engelbrecht1990}
J.~R. Engelbrecht and M.~Randeria, Phys.\ Rev.\ Lett.\ \textbf{65}, 1032 (1990).

\bibitem{engelbrecht1992}
J.~R. Engelbrecht and M.~Randeria, Phys.\ Rev.~B \textbf{45}, 12419 (1992).

\bibitem{ziman}
J.~M. Ziman, \emph{{Electrons and Phonons}} (Oxford University Press,
Oxford, 1960).

\bibitem{smith1989}
H.~Smith and H.~H. Jensen, \emph{{Transport phenomena}} (Oxford University
  Press, New York, 1989).

\bibitem{arnold}
P.~Arnold, G.~D. Moore, and L.~G. Yaffe, J. High Energy Phys.\ \textbf{11}, 001
  (2000).

\bibitem{mueller2011}
M.~M\"uller and H.~C. Nguyen, New J. Phys.\ \textbf{13}, 035009 (2011).

\bibitem{braby2011}
M.~Braby, J.~Chao, and T.~Sch{\"a}fer, New J. Phys.\ \textbf{13}, 035014 (2011).

\bibitem{taylor2012}
E.~Taylor and M.~Randeria, arXiv:1205.1525.

\end{thebibliography}
\end{document}